\documentclass[a4paper,11pt]{article}
\usepackage[table,xcdraw]{xcolor}
\usepackage{jinstpub} 
\graphicspath{{graphics}}

\usepackage{siunitx}
\usepackage{subcaption}

\usepackage{lineno}

\title{Pixel detector hybridisation with Anisotropic Conductive Films}

\author[a,b,1]{J.V. Schmidt\note{Corresponding author.}}
\author[a,c]{J. Braach}
\author[a]{D. Dannheim}
\author[a]{R. De Oliveira}
\author[a]{P. Svihra}
\author[d]{M. Vicente Barreto Pinto}

\affiliation[a]{CERN, EP-DT\\Meyrin, Switzerland}
\affiliation[b]{Karlsruhe Institute of Technology, Department of Mechanical Engineering\\Karlsruhe, Germany}
\affiliation[c]{University of Hamburg,  Institute of Experimental Physics\\Hamburg, Germany}
\affiliation[d]{University of Geneva, Départment de physique nucléaire et corpusculaire, \\Geneva, Switzerland}

\emailAdd{janis.viktor.schmidt@cern.ch}

\abstract{
Hybrid pixel detectors require a reliable and cost-effective interconnect technology adapted to the pitch and die sizes of the respective applications. During the ASIC and sensor R\&D phase, and in general for small-scale applications, such interconnect technologies need to be suitable for the assembly of single-dies, typically available from Multi-Project-Wafer submissions. Within the CERN EP R\&D programme and the AIDAinnova collaboration, innovative hybridisation concepts targeting vertex-detector applications at future colliders are under development. This contribution presents recent results of a newly developed in-house single-die interconnection process based on Anisotropic Conductive Film (ACF). The ACF interconnect technology replaces the solder bumps with conductive particles embedded in an adhesive film. The electro-mechanical connection between the sensor and the read-out chip is achieved via thermo-compression of the ACF using a flip-chip device bonder. A specific pad topology is required to enable the connection via conductive particles and create cavities into which excess epoxy can flow. This pixel-pad topology is achieved with an in-house Electroless Nickel Immersion Gold (ENIG) plating process that is also under development within the project. The ENIG and ACF processes are qualified with the Timepix3 ASIC and sensors, with \SI{55}{\micro\meter} pixel pitch and \SI{14}{\micro\meter} pad diameter. The ACF technology can also be used for ASIC-PCB/FPC integration, replacing wire bonding or large-pitch solder bumping techniques. This contribution introduces the ENIG plating and ACF processes and presents recent results on Timepix3 hybrid assemblies.
}


\proceeding{Topical Workshop on Electronics for Particle Physics - TWEPP2022\\
  19 - 23 September 2022\\
  Bergen, Norway}

\begin{document}
\maketitle
\flushbottom

\section{Introduction and motivation}\label{sec:intro}
Advanced fine-pitch interconnection processes suitable for single-die processing are under development within the CERN EP R\&D program on technologies for future experiments \cite{EPreport} and the AIDAinnova consortium \cite{AIDA}.\\
Small-pitch chip-to-chip interconnection is usually achieved in a sophisticated commercial bump-bonding process, requiring an Under-Bump Metallization (UBM) and bump deposition on wafer level. 
Therefore, this interconnect technology is not available as an affordable option for single-die bonding. Furthermore, there are long turn-over times, since the process has to be adapted to the specific wafer design. This is especially a problem for research and development (R\&D) or small volume projects, when the chips are manufactured on Multi-Project-Wafers (MPW).\\
An alternative of interconnection process using Anisotropic Conducting Film (ACF) is currently under development, which replaces the solder-bump connection by thermo-compression of Conductive Particles (CP) embedded in a thin adhesive film \cite{Svihra2022}. The needed pad topology, to compress the CPs, is achieved with a mask-less Electroless Nickel Immersion Gold (ENIG) plating, adapted to the respective pad geometry. 
The ACF technology has already been used in the mass production of LCD displaces for several decades \cite{Myung-Jin1998} and is increasingly used for chip-to-film or chip-to-flex connections \cite{Chang2001}. Efforts have been made to adapt the commercial process to detector integration \cite{Svihra2022}.  
\\Initial proof-of-concept studies described in the following have been performed using Timepix3 hybrid ASICs \cite{Timepix3} and matching sensors.

\section{Electroless Nickel Immersion Gold (ENIG) plating}\label{sec:ENIG}
To obtain the required topology for the ACF bonding, a single-die in-house Electroless Nickel Immersion Gold (ENIG) plating process is developed as part of this project. 
First, the aluminium oxide layer is removed from the pads. Then, the pad surface is activated to create a catalytic surface on which the reaction can start in the subsequent electroless nickel bath. This is achieved by the commonly used zincation step \cite{Arshad2004}. In the hypophosphite-based electroless nickel bath, the zincated pads act as catalytic surfaces and start the reaction. 
The reaction of the nickel deposition can be written as: 
\begin{equation}
    \mathrm{2 H_2PO_2^- + Ni^{2+} + 2 H_2O \rightarrow 2H_2PO_3^- + H_2 + 2 H^+ + Ni^0}
    \label{equ:Ni-sum}
\end{equation}

\begin{figure}
    \centering
    \includegraphics[width=1.\textwidth]{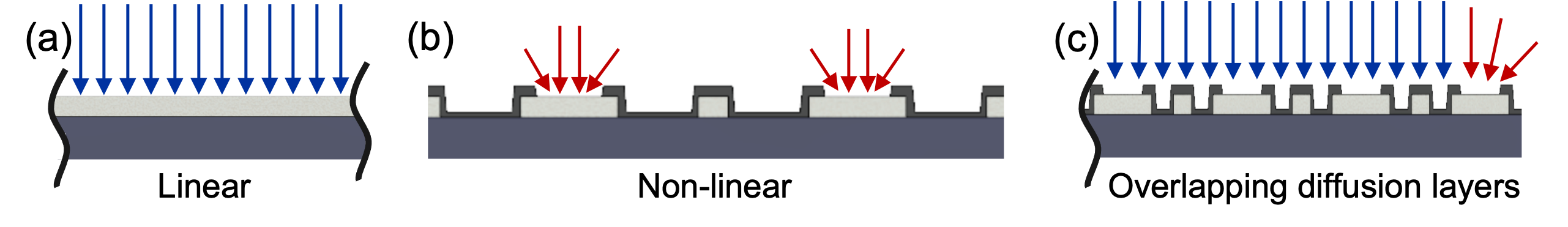}
    \caption{Illustration of diffusion to different structured surfaces, with linear (blue arrows) and non-linear (red arrows) diffusion.}
    \label{fig:diffmech}
\end{figure}

The plated nickel layers on the zincated pads also act as catalytic surfaces \cite{Arshad2004}, moreover the amount of the reacted chemicals is negligibly small in relation to the total amount in the solution, therefore the autocatalytic reaction continues until the sample is removed from the bath. Thus, the plating height is controlled by the reaction time.\\
The last step is the immersion gold bath for corrosion protection. In the bath, the outer layer of nickel atoms are replaced by gold atoms, the equation can be written as: 
\begin{equation}
    \mathrm{Ni + 2 Au^+ \rightarrow Ni^{2+} + 2 Au}
    \label{equ:IG}
\end{equation}
\begin{figure}
    \centering
    \includegraphics[width=0.8\textwidth]{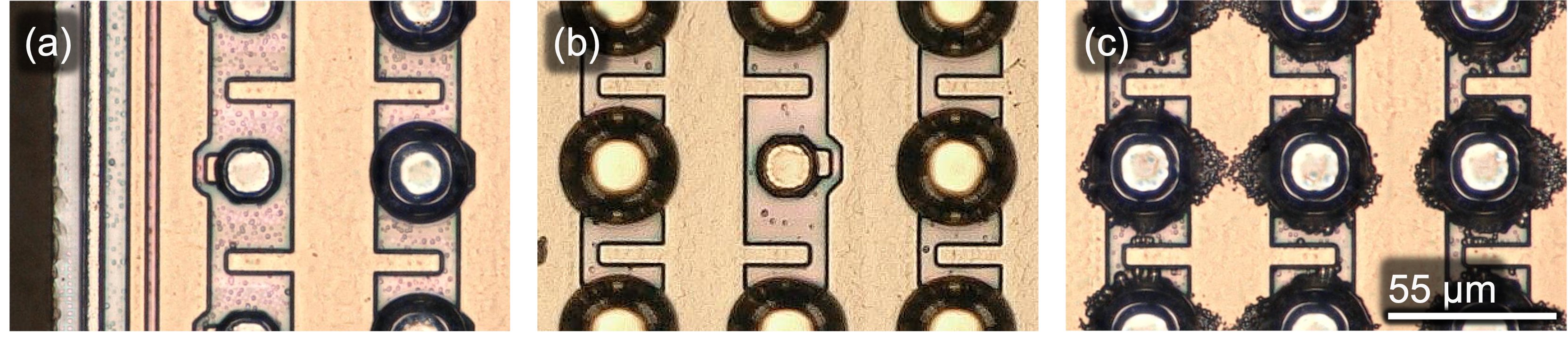}
    \caption{Observed defective plating on Timepix3 ASICs (a) missing plating next to edge, (b) skip plating and (c) overplating.}
    \label{fig:plating_def}
    \label{fig:defplating}
\end{figure}
During the process adjustment for the Timepix3 ASIC, three types of defective plating have been observed repeatedly (Figure~\ref{fig:defplating}).
Due to the autocatalytic characteristic of the plating, the reaction can also start a self-accelerating chain reaction on surfaces of solid particles or colloids in the solution. To prevent this degradation, stabilizers acting as a catalyst poison are used in the nickel baths. However, they are also adsorbed on the pad surfaces.
The adsorbed stabilizer is continuously buried under freshly plated nickel, and the reaction only stops if the stabilizer adsorption is faster than the burying rate \cite{Zhang1999}, which is proportional to the reaction speed of the nickel deposition (depending on pH value, temperature and hypophosphite concentration) \cite{Siau2006}. The adsorption is proportional to the diffusion flux of the stabilizer and is strongly influenced by the diffusion mechanism and pad size \cite{Zhang1999}. If the pad diameter is several hundred \SI{}{\micro\meter}, the edge effect is negligibly small and the diffusion can be considered linear (Figure~\ref{fig:diffmech}a).
However, if the pads are smaller than the diffusion layer, the edge effect (non-linear diffusion) will prevail (Figure~\ref{fig:diffmech}b). If the diffusion layers
of several small pads overlap, the diffusion for all but the edge pads can be considered as linear again \cite{Zhang1999}. The described different mechanisms are illustrated in Figure~\ref{fig:diffmech}. 
Therefore, the missing plating near the edge (Figure~\ref{fig:plating_def}a) and skip plating (Figure~\ref{fig:plating_def}b) are assumed to be caused by catalytic poisoning of the pad surfaces \cite{Zhang1999,Siau2006}. \\
On the other hand, if the electroless nickel solution gets too active by decreasing the stabilizer concentration or increasing the reaction speed, plating also happens on unwanted places like the passivation \cite{Siau2006}. This overplating was also observed on the Timepix3 ASICs, as seen in Figure~\ref{fig:plating_def}c.\\

\section{Bonding with ACFs}
\begin{figure}[h!]
    \centering
    \includegraphics[width=0.85\textwidth]{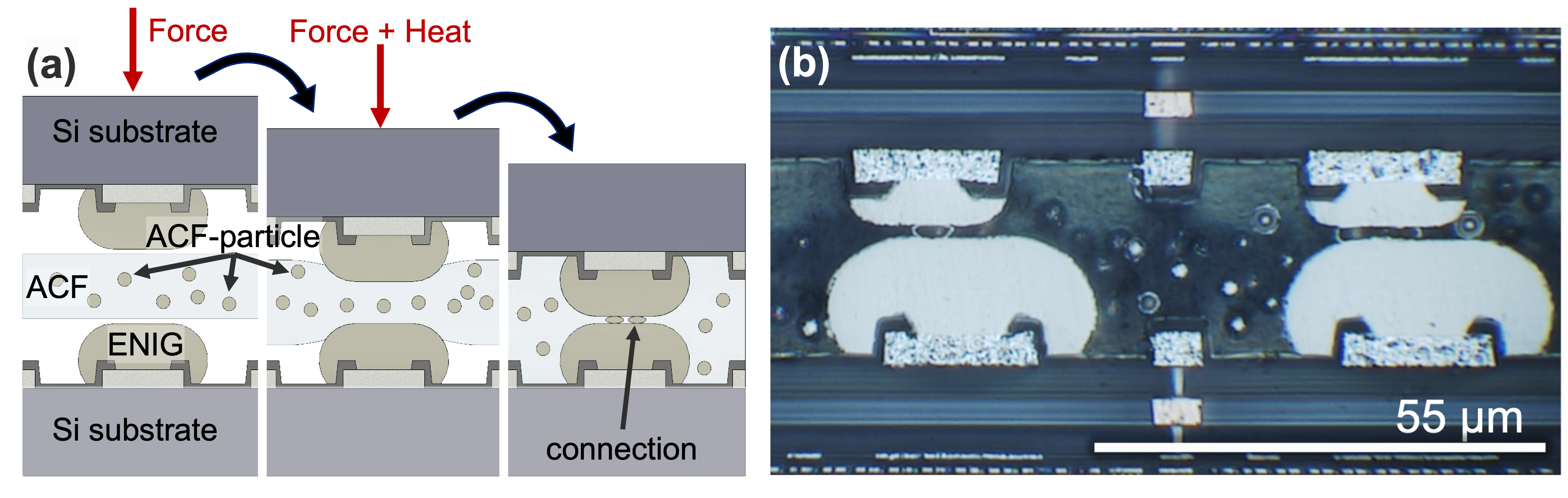}
    \caption{(a) Illustration of the ACF bonding steps and (b) microscope image of a cross-section of a dummy sample.}
    \label{fig:bonding}
\end{figure}

\begin{figure}[t]
    \centering
    \includegraphics[width=0.8\textwidth]{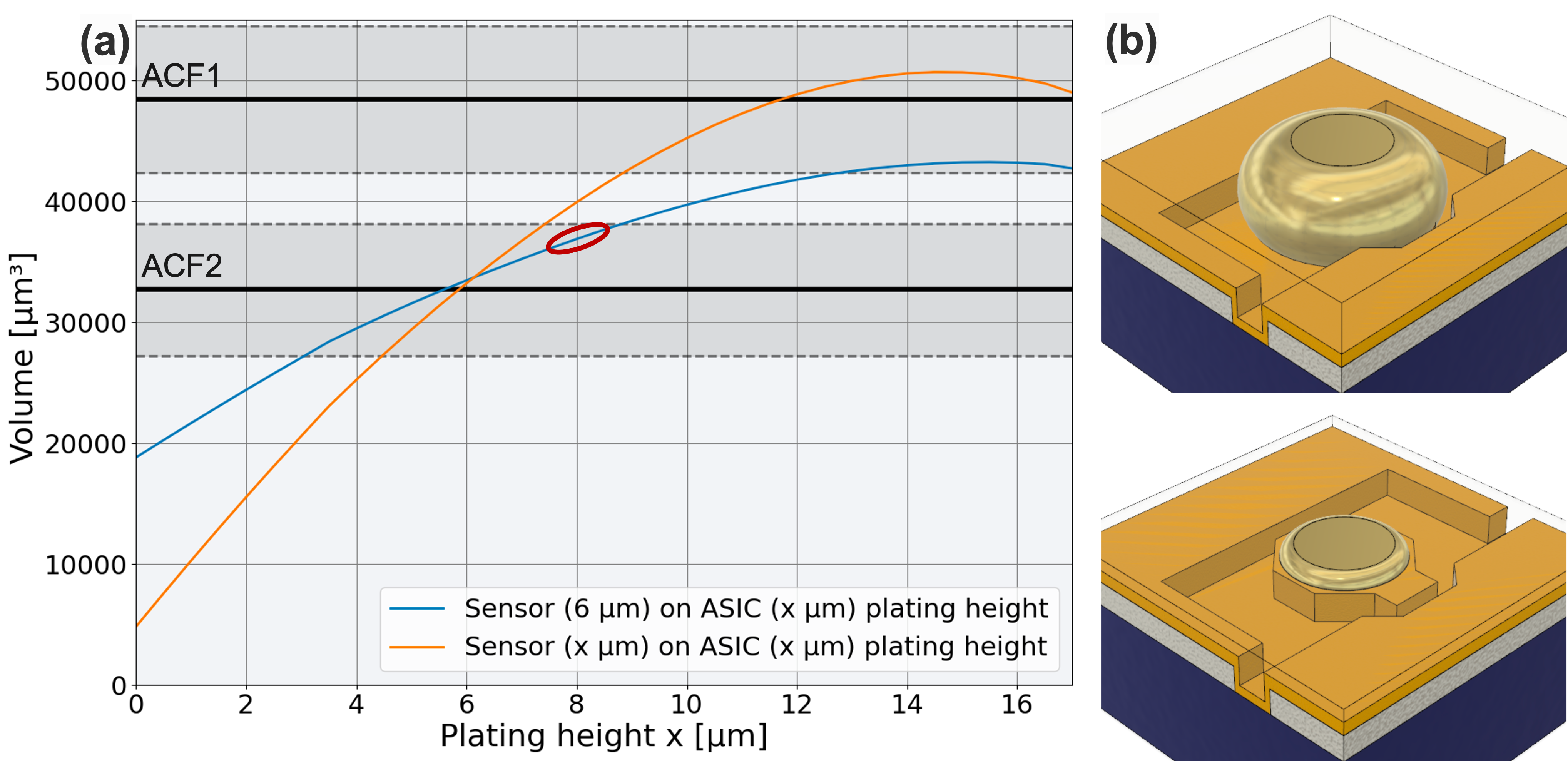}
    \caption{(a) Volume of the cavities as a function of plating height. The red ellipse marks the cavity volume of the two samples S1 and S2 from Figure~\ref{fig:hitmap}.  (b) Visualization of two Timepix3 pixels with different ENIG heights.}
    \label{fig:cavity}
\end{figure}

As part of the bonding step, the electromechanical connection is created by compressing the CPs between the pads and by heat-curing the resin, which is illustrated in Figure~\ref{fig:bonding}a. The required pressure and temperature depend on the used ACF. The flip-chip bonding machine used has a maximal force of \SI{980}{\newton}, and can achieve up to \SI{400}{\celsius}. Although, the ACFs used cure at \SI{150}{\celsius} in \SI{5}{\second}. \\
To fully connect a Timepix3 ASIC a total of $256\times256$ pixels on a size of $14.1\times\SI{14.1}{\milli\meter\squared}$  with a \SI{55}{\micro\meter} pixel pitch have to be bonded. This is a crucial difference to the standard application, in which the connected area is thin and long, typically only \SI{2}{\milli\meter} wide and several cm long. \\
For the trials, two ACF samples with \SI{3}{\micro\meter} CPs were used, referred to as ACF1 and ACF2.
ACF1 and ACF2 have a thickness, particle density and bonding pressure of \SI{18}{\micro\meter}, 71,000 CP/mm$^2$, 30~--~\SI{80}{\mega\pascal} and \SI{14}{\micro\meter}, 60,000 CP/mm$^2$, 50~--~\SI{90}{\mega\pascal}, respectively.
For ACF2 an uncertainty in thickness of \SI{\pm2}{\micro\meter} was given and for simplicity, this was also assumed for ACF1. \\ 
During the bonding process, the pads come close together and excess adhesive has to be displaced. As the distance to the edge increases, it becomes more difficult to force the excess epoxy out to the sides and the volume of the cavities between the pads have to fit all the excess epoxy. 
If the cavities are not sufficently spacious, only the area near the edge will be connected.
The volume of the cavities is directly related to the plating height, as shown in the Figure~\ref{fig:cavity}.
An approximate model was developed that relates the cavity volume to the ENIG plating height (Figure~\ref{fig:cavity}a). It was made to evaluate the results from different ACF materials and obtain the target plating height for future samples.\\  
For the ACFs it was assumed that the particles have to be compressed by about 30\% to achieve a good connection, which results in a persistent gap of \SI{2}{\micro\meter} for \SI{3}{\micro\meter} CPs, that was taken into account in the model. For the calculation of the volume of ACF2 a factor of 0.9 was added, since it is supplied and applied in \SI{2}{\milli\meter} wide stripes and, after taking the gaps into account, the total coverage of the matrix was about 90\%.\\
With this assumption, a volume of $(16\pm2)\times55\times\SI{55}{\micro\meter\cubed}$ (\SI{42,350}{\micro\meter\cubed}~--~\SI{54,450}{\micro\meter\cubed}) for ACF1 and $(0.9\times12\pm2)\times55\times\SI{55}{\micro\meter\cubed}$ (\SI{27,225}{\micro\meter\cubed}~--~\SI{38,115}{\micro\meter\cubed}) for ACF2 have to fit in the cavities between the pads as indicated in the graph (Figure~\ref{fig:cavity}a). 

\section{Results}\label{sec:result}
\begin{figure}
    \centering
    \includegraphics[width=0.9\textwidth]{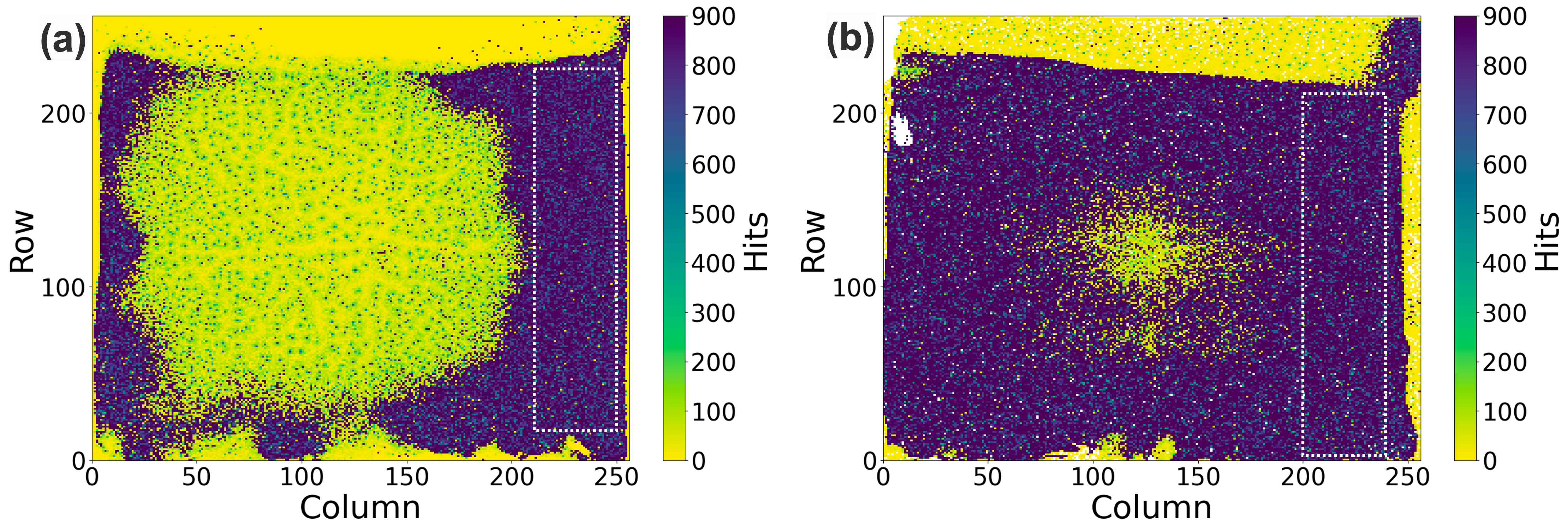}
    \caption{Hit-maps from a Sr$^{90}$ illumination of the Timepix3 samples (a) S1 and (b) S2. The white dashed  framed areas are used for the estimation of the connection yield. }
    \label{fig:hitmap}
\end{figure}

For process development, dummy samples were processed and then cut to reveal cross-sections for the examination of the pad distance and alignment, as seen in Figure~\ref{fig:bonding}b. \\
With electrically functioning samples, source measurements were carried out and are discussed for two samples, referred to as S1 and S2, bonded using ACF1 and ACF2, respectively. Both samples underwent the same treatment before bonding and therefore have the same plating height of roughly \SI{8}{\micro\meter} on the Timepix3 ASIC and \SI{6}{\micro\meter} on the sensor.  
The resulting hit maps from a Sr$^{90}$ source exposure for the two bonded assemblies are shown in Figure~\ref{fig:hitmap}.
The missing connection next to the edge is caused by defective plating, as discussed in section \ref{sec:ENIG}. The middle of both samples contains a circular spot of missing or badly connected pixels, which cannot be explained by defective plating. The affected area is larger for the thicker ACF. Therefore, it is assumed that the difference is related to the adhesive volume, as shown in Figure~\ref{fig:cavity}a. The cavity volume of both samples of about \SI{35,000}{\micro\meter\cubed}, marked with the red eclipse in Figure~\ref{fig:cavity}a, is too small to fit the volume of the \SI{18}{\micro\meter} thick ACF1. Even for the thinner ACF2, the cavity volume is not sufficient after taking the uncertainty on the thickness into account. 

\subsection*{Estimate of connection yield}
\begin{figure}
    \centering
    \includegraphics[width=0.9\textwidth]{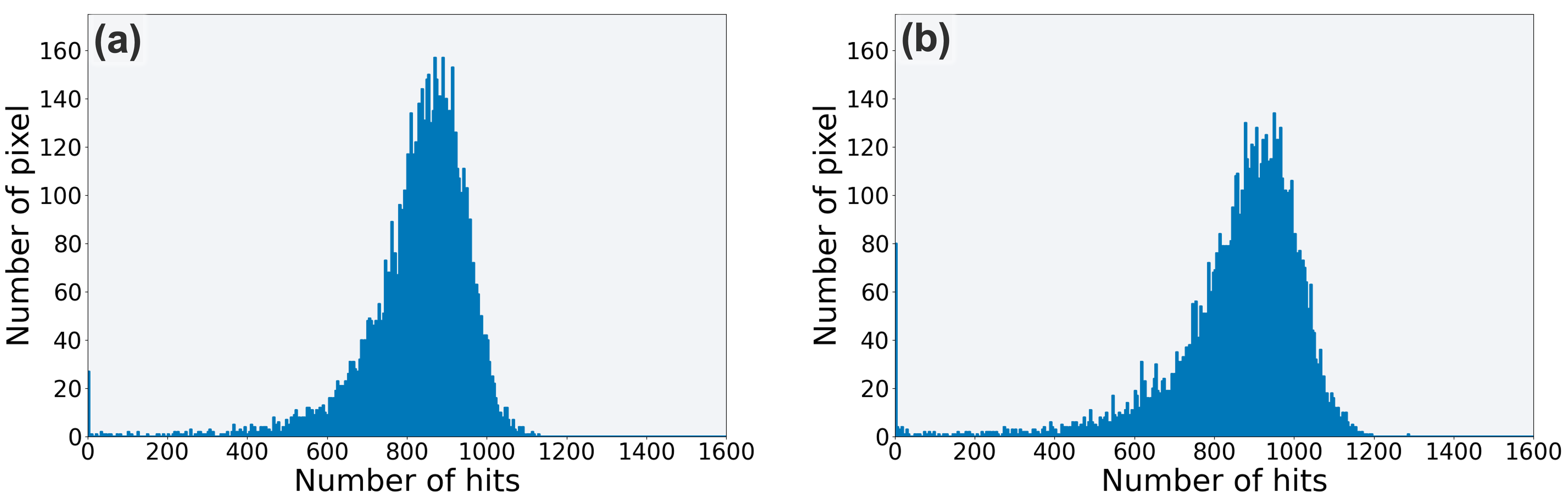}
    \caption{Histograms of (a) S1 and (b) S2 of pixels in the well-connected areas (Figure~\ref{fig:hitmap}).}
    \label{fig:histogram}
\end{figure}

By analysing well-connected areas, the response of boded pixels can be evaluated in more detail. An area of $40\times210$  pixels on the right part of the hit-maps of both samples was used, as indicated with the white frames in (Figure~\ref{fig:hitmap}). Of the 8400 pixel investigated, the samples S1 and S2 have 27 and 72 pixels with zero hits, respectively. 
The indicated pixel hits are also evaluated as histograms (Figure~\ref{fig:histogram}), which show a high connection yield.  In the histogram, a tail to lower hit counts becomes visible, which will be investigated in-depth with future measurements. 
\section{Summary and outlook}
Proof-of-concept hybridisation tests using thermo-compression of ACFs for Timepix3 ASICs and sensors, plated with a new in-house ENIG process, have demonstrated the potential of the process for bonding pixel detectors. The ENIG plating of single-dies with small pads revealed to be a challenge and is still undergoing improvements. The cavity volume between the pads, which is related to the ENIG height, was identified as a crucial process parameter for successful area ACF bonding of small pitch hybrid detectors.\\
High connection yields have been observed in well-connected areas. In order to calculate accurate yield values and analyse possible failure types, beam tests and more laboratory measurements will be carried out with existing and new samples. 

\acknowledgments
This project has received funding from the European Union’s Horizon 2020 Research and Innovation programme under GA no 101004761.

\end{document}